# Quantum deletion: Beyond the no-deletion principle

Satyabrata Adhikari
Department of Mathematics,
Bengal Engineering College & Science University
Howrah-711103, West Bengal, India.
E-Mail: satyyabrata@yahoo.com

**Abstract:** Suppose we are given two identical copies of an unknown quantum state and we wish to delete one copy from among the given two copies. The quantum no-deletion principle restricts us from perfectly deleting a copy but it does not prohibit us from deleting a copy approximately. Here we construct two types of a " universal quantum deletion machine" which approximately deletes a copy such that the fidelity of deletion does not depend on the input state. The two types of universal quantum deletion machines are (1) a conventional deletion machine described by one unitary operator and (2) a modified deletion machine described by two unitary operators. Here it is shown that modified deletion machine deletes a qubit with fidelity $3/4$, which is the maximum limit for deleting an unknown quantum state. In addition to this we also show that the modified deletion machine retains the qubit in the first mode with average fidelity 0.77 (approx.) which is slightly greater than the fidelity of measurement for two given identical state, showing how precisely one can determine its state [13]. We also show that the deletion machine itself is input state independent i.e. the information is not hidden in the deleting machine, and hence we can delete the information completely from the deletion machine.

**I. Introduction:**

The impossibility of perfect cloning [1] and perfect deletion [7] are two fundamental laws that nature imposes on quantum information theory. Although the linearity of quantum theory prohibit us from duplicating and deleting an unknown quantum state accurately it cannot forbid us from constructing an approximate cloning [1- 4] and deletion machine [7,8,11,14]. The possibility of perfect copying and perfect deletion with probability less than 1 [5,6,9,10] is now well-established fact. Buzek et.al. [2] showed the existence of a "universal quantum copying machine" which approximately copies quantum-mechanical states such that the quality of its output does not depend on the input. The universal cloning transformation was shown to be optimal by Gisin et.al. [3]. Like the universal quantum cloning machine, D.Qiu [11] constructed a universal deletion machine but unfortunately the machine is found to be non-optimal in the sense of low fidelity of deletion. Recently in [14] we studied the deletion of one copy of a qubit from two imperfect cloned copies obtained from Buzek-Hillery and Wotters-Zurek quantum cloning machines separately.

In this work, we constructed two types of deletion machine. The first type of deletion machine is conventional, i.e., it just deletes a qubit, while the second type of deletion machine not only deletes a qubit but also transforms the state after deletion operation. Later we will find that the conventional deletion machine deletes a qubit with fidelity $1/2$ for all input state while the modified deletion machine (second type) deletes a qubit of arbitrary input state with the same fidelity $3/4$, which is the maximum limit for deleting an unknown qubit [13].

The newly defined deletion machine i.e. the modified deletion machine, consists of two parts, the deleter and the transformer.

**1. Deleter:** This is nothing but a unitary transformation U used to delete one copy from among two given copies of an unknown quantum state.

A unitary transformation U which describes a deleter is given below:

$$U|00\rangle_{ab}|A\rangle_c \rightarrow |0\rangle_a|\Sigma\rangle_b|A_0\rangle_c + \left[|0\rangle_a|1\rangle_b + |1\rangle_a|0\rangle_b\right]|B_0\rangle_c \tag{1}$$

$$U|01\rangle_{ab}|A\rangle_c \rightarrow |0\rangle_a|\Sigma_\perp\rangle_b|D_0\rangle_c + |1\rangle_a|0\rangle_b|C_0\rangle_c \tag{2}$$

$$U|10\rangle_{ab}|A\rangle_c \rightarrow |1\rangle_a|\Sigma\rangle_b|D_0\rangle_c + |0\rangle_a|1\rangle_b|C_0\rangle_c \tag{3}$$

$$U|11\rangle_{ab}|A\rangle_c \rightarrow |1\rangle_a|\Sigma_\perp\rangle_b|A_1\rangle_c + \left[|0\rangle_a|1\rangle_b + |1\rangle_a|0\rangle_b\right]|B_1\rangle_c \tag{4}$$

where $|A\rangle$ is the initial and $|A_i\rangle, |B_i\rangle, |C_j\rangle, |D_j\rangle$ ($i=0,1; j=0$) are the final machine state vectors. $|\Sigma\rangle$ is some standard state and $|\Sigma_\perp\rangle$ denote a state orthogonal to $|\Sigma\rangle$.

The deleter used in the deletion machine deletes a qubit irrespective of the input qubits, which can be identical or different. In addition, we note that, when $|0\rangle$ occur in the second mode, the deletion machine transform the qubit to the state $|\Sigma\rangle$ while $|1\rangle$ (orthogonal to $|0\rangle$) in the second mode is transformed into $|\Sigma_\perp\rangle$.

We assume

$$\langle A_0|B_0\rangle = \langle A_0|D_0\rangle = \langle A_1|D_0\rangle = \langle A_1|B_1\rangle = \langle A_0|A_1\rangle = \langle B_0|C_0\rangle = \langle B_0|D_0\rangle = 0. \quad (5)$$

$$\langle A_0|B_1\rangle = \langle A_1|B_0\rangle = \langle B_0|B_1\rangle = \langle B_1|D_0\rangle = \langle C_0|A_1\rangle = \langle B_1|C_0\rangle = 0.$$

$$\langle A|A_0\rangle = \langle A|D_0\rangle = \langle A|A_1\rangle = Y, \langle A|B_0\rangle = \langle A|C_0\rangle = \langle A|B_1\rangle = 0 \quad (6)$$

The normalization condition of the transformation (1-4) gives

$$\langle A_i|A_i\rangle + 2\langle B_i|B_i\rangle = 1, \; i = 0,1$$
$$\langle C_0|C_0\rangle + \langle D_0|D_0\rangle = 1 \quad (7)$$

The orthogonality condition to be satisfied for the transformation (1-4) is

$$\langle A_0|C_0\rangle = \langle D_0|C_0\rangle = 0 \quad (8)$$

**2. Transformer:** It is described by a unitary transformation T. It is used in the deletion machine in such a way to increase the fidelity of deletion and minimize the distortion of the undeleted qubit.

The unitary operator T [12] is defined by

$$T = |\psi^+\rangle\langle 00| + |11\rangle\langle 01| + |\psi^-\rangle\langle 10| + |00\rangle\langle 11| \quad (9)$$

Where $|\psi^\pm\rangle = (1/\sqrt{2})(|01\rangle \pm |10\rangle)$

**Few Definitions:**

Let $|\psi\rangle = \alpha|0\rangle + \beta|1\rangle$ with $\alpha^2 + |\beta|^2 = 1$ be any unknown quantum state.

Without any loss of generality we can take $\alpha$ real and $\beta$ complex.

Let $\rho_a$ and $\rho_b$ be the reduced density operator describing the state of the undeleted qubit in mode a and the state of the deleted qubit in mode b, respectively, and $\rho_c$ denotes the density operator of the machine state after the deleting operation.

Let $F_a = \langle\psi|\rho_a|\psi\rangle$, $F_b = \langle\Sigma'|\rho_b|\Sigma'\rangle$, where $|\Sigma'\rangle = (1/\sqrt{2})(|\Sigma\rangle + |\Sigma_\perp\rangle)$, denotes the fidelity of the qubit in the modes a and b, respectively, after the deletion operation and $F_c = \langle A|\rho_c|A\rangle$ denotes the overlapping between the initial and final machine state vectors.

**Definition 1: State dependent deletion machine**

A deletion machine is said to be state dependent if $F_a, F_b$ and $F_c$ depends on the input state.

**Definition 2: Universal deletion machine**

A deletion machine is said to be universal if $F_b$ and $F_c$ are independent of the input state. The machine is optimal if it maximizes $F_a$ and $F_b$, respectively.

**Definition 3: Ideal deletion machine**

A deletion machine is said to be ideal if $F_a$, $F_b$ and $F_c$ are input state independent and the machine is optimal if it maximizes $F_a$ and $F_b$, respectively.

**Note:** From definition 2 and 3, we can say that every ideal deletion machine is universal but converse is not true.

## II. Conventional deletion machine (Deletion machine without transformer):

In this section, we aim to show that the conventional deletion machine (1- 4) i.e. deletion machine without transformer, becomes either universal deletion machine or an ideal deletion machine in some restricted cases.

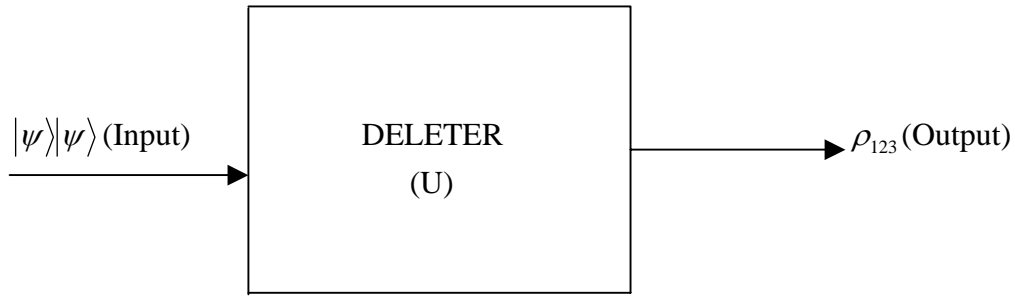

Conventional deletion machine

Fig.1 When an input state $|\psi\rangle|\psi\rangle$ is given into the deletion machine to delete a qubit, the deletion machine deletes a qubit and $\rho_{123}$ describes the output state of the deletion machine. 1, 2, 3 denotes the undeleted, deleted and machine state modes, respectively.

The deletion machine can be shown to be universal or ideal following three steps.

**Step 1:** The reduced density operator in the mode 1 is given by

$$\rho_1 = Tr_{23}(\rho_{123}) = |0\rangle\langle 0| \left[\alpha^4(\langle A_0|A_0\rangle + \langle B_0|B_0\rangle) + \alpha^2|\beta|^2 + |\beta|^4\langle B_1|B_1\rangle\right] \\ + |1\rangle\langle 1| \left[\alpha^4\langle B_0|B_0\rangle + \alpha^2|\beta|^2 + |\beta|^4(\langle A_1|A_1\rangle + \langle B_1|B_1\rangle)\right] \quad (10)$$

Let us assume $\langle A_0|A_0\rangle = \langle A_1|A_1\rangle = \langle D_0|D_0\rangle = 1-2\lambda$ (11)

$$and \ \langle B_0|B_0\rangle = \langle B_1|B_1\rangle = \langle C_0|C_0\rangle/2 = \lambda \qquad (12)$$

with $0 \leq \lambda \leq 1/2$ which follows from Schwarz inequality.

$$\rho_1 = |0\rangle\langle 0|\left[\alpha^4(1-\lambda)+\alpha^2|\beta|^2+|\beta|^4\lambda\right]+|1\rangle\langle 1|\left[\alpha^4\lambda+\alpha^2|\beta|^2+|\beta|^4(1-\lambda)\right] \qquad (13)$$

The overlapping between the input state $|\psi\rangle$ and the density operator $\rho_1$ is given by

$$F_1 = \langle\psi|\rho_1|\psi\rangle = (1-\lambda)+2\alpha^2|\beta|^2(2\lambda-1) = (1-\lambda)+2\alpha^2(1-\alpha^2)(2\lambda-1) \qquad (14)$$

Therefore $F_1$ depends on $\alpha^2$ and the parameter $\lambda$.

Now it seems to be interesting to discuss results for two different values of $\lambda$ in two different cases. In the first case $F_1$ is found to be input state independent and in the second case it depends on the input state.

**Case I:** If $\lambda \to 1/2$ then $F_1 \to 1/2$. Although we are able to make $F_1$ input state independent, the performance of the deletion machine is not very satisfactory since it fails to retain the qubit in the first mode faithfully after the deletion operation.

**Case II:** If $\lambda \to 0$, then $F_1 \to 1-2\alpha^2|\beta|^2 = 1-2\alpha^2(1-\alpha^2)$, which is input state dependent and therefore we have to calculate the average value.

The average fidelity is given by

$$\overline{F_1} = \int_0^1 F_1(\alpha^2)d\alpha^2 \to 2/3. \qquad (15)$$

This value is equal to the fidelity of measurement, which is for a given single unknown state. Although $F_1$ is input state dependent but the average value $\overline{F_1}$ exceeds the fidelity discussed in case I.

**Step 2:** The reduced density operator in the mode '2' is given by

$$\begin{aligned}\rho_2 = Tr_{13}(\rho_{123}) = &|0\rangle\langle 0|\left[\alpha^4\langle B_0|B_0\rangle+\alpha^2|\beta|^2\langle C_0|C_0\rangle+|\beta|^4\langle B_1|B_1\rangle\right] \\ &+|1\rangle\langle 1|\left[\alpha^4\langle B_0|B_0\rangle+\alpha^2|\beta|^2\langle C_0|C_0\rangle+|\beta|^4\langle B_1|B_1\rangle\right] \\ &+|\Sigma\rangle\langle\Sigma|\left[\alpha^4\langle A_0|A_0\rangle+\alpha^2|\beta|^2\langle D_0|D_0\rangle\right] \\ &+|\Sigma_\perp\rangle\langle\Sigma_\perp|\left[|\beta|^4\langle A_1|A_1\rangle+\alpha^2|\beta|^2\langle D_0|D_0\rangle\right]\end{aligned} \qquad (16)$$

Using equations (11) & (12), equation (16) reduces to

$$\rho_2 = |0\rangle\langle 0| \left[\alpha^4\lambda + 2\alpha^2|\beta|^2\lambda + |\beta|^4\lambda\right] + |1\rangle\langle 1| \left[\alpha^4\lambda + 2\alpha^2|\beta|^2\lambda + |\beta|^4\lambda\right]$$
$$+ |\Sigma\rangle\langle\Sigma| \left[\alpha^2(1-2\lambda)\right] + |\Sigma_\perp\rangle\langle\Sigma_\perp| \left[|\beta|^2(1-2\lambda)\right] \quad (16A)$$

The fidelity of deletion is defined by

$$F_2 = \langle\Sigma'|\rho_2|\Sigma'\rangle = (1/2)\left[(1-2\lambda) + (K_1 + K_2)\lambda\right] \quad (17)$$

$$K_1 = \langle\Sigma|0\rangle^2 + |\langle\Sigma|1\rangle|^2 + \langle\Sigma|0\rangle\langle 0|\Sigma_\perp\rangle + \langle\Sigma|1\rangle\langle 1|\Sigma_\perp\rangle \quad (18)$$

$$K_2 = |\langle\Sigma_\perp|0\rangle|^2 + \langle\Sigma_\perp|1\rangle^2 + \langle\Sigma|0\rangle\langle 0|\Sigma_\perp\rangle + \langle\Sigma|1\rangle\langle 1|\Sigma_\perp\rangle \quad (19)$$

The standard state $|\Sigma\rangle$ can be written as $|\Sigma\rangle = m_1|0\rangle + m_2|1\rangle$, where without any loss of generality we can take $m_1$ real and $m_2$ complex quantities and satisfies the relation

$$m_1^2 + |m_2|^2 = 1. \quad (20)$$

A state orthogonal to $|\Sigma\rangle$ is given by $|\Sigma_\perp\rangle = -m_2^*|0\rangle + m_1|1\rangle$. \quad (21)

Therefore $\langle\Sigma|0\rangle = \langle\Sigma_\perp|1\rangle = m_1$, $\langle\Sigma|1\rangle = m_2^*$, $\langle\Sigma_\perp|0\rangle = -m_2$ \quad (22)

Using equations (18), (19), (20) & (22), we get $K_1 + K_2 = 2$. \quad (23)

Putting the value of $K_1 + K_2$ in equation (17), we get

$$F_2 = 1/2 \quad (24)$$

Here we note that the fidelity of deletion depends on neither the input state nor the machine state but is calculated to be $1/2$, which is not a very satisfactory result at all. The same fidelity is also obtained by D.Qiu [11] for his deletion machine and it emphasizes the difficulty for improving its fidelity. We also find here that the fidelity of deletion for our prescribed deletion machine cannot be improved further if the machine is kept in its present form but the fidelity may be improved if we define a deletion machine in another way, which we discuss in detail in the next section.

**Step 3:** The reduced density operator in the mode '3' is given by

$$\rho_3 = Tr_{12}(\rho_{123}) = \alpha^4 \{|A_0\rangle\langle A_0| + m_2^*|A_0\rangle\langle B_0| + m_2|B_0\rangle\langle A_0| + 2|B_0\rangle\langle B_0|\} +$$

$$\alpha^3\beta^* \{m_2|A_0\rangle\langle C_0| + 2m_1|B_0\rangle\langle D_0| + 2|B_0\rangle\langle C_0|\} +$$

$$\alpha^3\beta \{m_2^*|C_0\rangle\langle A_0| + 2m_1|D_0\rangle\langle B_0| + 2|C_0\rangle\langle B_0|\} +$$

$$\alpha^2|\beta|^2 \begin{Bmatrix} m_2\langle A_0|B_1\rangle + m_2^*\langle B_1|A_0\rangle - m_2^*|A_1\rangle\langle B_0| - m_2|B_0\rangle\langle A_1| + 2(\langle B_0|B_1\rangle + \langle B_1|B_0\rangle) + \\ 2(|C_0\rangle\langle C_0| + |D_0\rangle\langle D_0|) + 2m_1(|C_0\rangle\langle D_0| + |D_0\rangle\langle C_0|) \end{Bmatrix} +$$

$$\alpha|\beta|^2\beta\{2m_1(|B_1\rangle\langle D_0| + |D_0\rangle\langle B_1|) - m_2^*|A_1\rangle\langle C_0| - m_2|C_0\rangle\langle A_1| + 2(|B_1\rangle\langle C_0| + |C_0\rangle\langle B_1|)\} +$$

$$|\beta|^4\{|A_1\rangle\langle A_1| - m_2^*|A_1\rangle\langle B_1| - m_2|B_1\rangle\langle A_1| + 2|B_1\rangle\langle B_1|\} \qquad (25)$$

Using equation (6) and the relation $\alpha^2 + |\beta|^2 = 1$, we get $\langle A|\rho_3|A\rangle = Y^2$ \qquad (26)

which is independent of $\alpha^2$. This means that the information is not hidden in the deletion machine and hence it deletes the state completely because we cannot retrieve the state by applying unitary transformation from the deletion machine

**Note:** (1) If $\lambda \to 1/2$, then $F_1$, $F_2$ and $\langle A|\rho_3|A\rangle$ are independent of $\alpha^2$. Also $F_1 \to 1/2$, $F_2 = 1/2$. Therefore, for $\lambda \to 1/2$ the conventional deletion machine becomes an ideal deletion machine but the machine is not optimal.

(2) If $\lambda \neq 1/2$ then also $F_2$ and $\langle A|\rho_3|A\rangle$ are independent of $\alpha^2$ because they do not depends on $\lambda$ so the conventional deletion machine becomes a universal deletion machine for all values of $\lambda$ $(0 \leq \lambda < 1/2)$.

Now case II is interesting in the sense that if $\lambda \to 0$, then the average value of $F_1$ tends to the maximum limit $2/3$ that is also obtained by state dependent Pati-Braunstein deletion machine. Moreover, the fidelity of a qubit in mode 1, i.e., $F_1$, is found to be greater than the fidelity of deletion $F_2$.

### III. Modified deletion machine (Deletion machine with transformer):

In the preceding section, we discuss the deletion machine without considering a vital part of it. In this section we take into account that important part of the deletion machine without which we cannot improve the fidelity of deletion. In addition to a unitary transformation U (named the deleter) that deletes a qubit, an unitary operator T (named the transformer) must be used in the deletion machine. The role of the transformer is to transform the resultant state immediately obtained after the deletion operation thereby improving the fidelity of deletion of the qubit in the second mode and increasing the fidelity of the retained qubit in the first mode.

Our protocol is described in the figure given below:

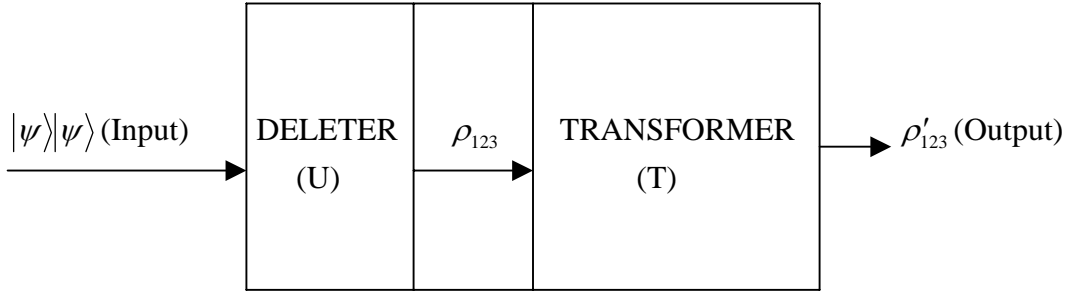

Modified Deletion Machine

Fig.2 The modified deletion machine consists of two chambers. The first chamber contains deleter described by the unitary operator U. It deletes one copy from among two copies $|\psi\rangle|\psi\rangle$. After deletion operation, the resultant state described by the density operator $\rho_{123}$ is allowed to pass through the transformer. The transformer is nothing but a unitary transformation T. The transformer T transforms the state $\rho_{123}$ to the state $\rho'_{123}$, which describes the output state of the deletion machine.

After the deletion operation (1- 4), the output of the deleter is described by the density operator $\rho_{123}$. Then the transformer used in the deletion machine transforms the state described by the density operator $\rho_{123}$ to the state $\rho'_{123} = (I \otimes T)\rho_{123}(I \otimes T)^{t}.$ (27)

The reduced density operator describing the state $\rho'_1$ is given by

$$\rho'_1 = |0\rangle\langle 0| \ (1/2) \begin{bmatrix} \alpha^4 \{m_1^2(1-2\lambda)+\lambda\} + \alpha^2|\beta|^2 \{(3|m_2|^2 - m_1(m_2+m_2^*) + m_1^2)(1-2\lambda) + 2\lambda\} + \\ |\beta|^4 \{(|m_2|^2 + 2m_1^2)(1-2\lambda) + \lambda\} \end{bmatrix}$$

$$+ |0\rangle\langle 1| \ (1/\sqrt{2}) \begin{bmatrix} \alpha^4 \{m_1 m_2^*(1-2\lambda) + \lambda\} + \alpha^2|\beta|^2 \{2\lambda + (m_1^2 - m_2^2 - m_1(m_2+m_2^*))(1-2\lambda)\} + \\ |\beta|^4 \{\lambda + m_1 m_2(1-2\lambda)\} \end{bmatrix}$$

$$+ |1\rangle\langle 0| \ (1/\sqrt{2}) \begin{bmatrix} \alpha^4 \{m_1 m_2(1-2\lambda) + \lambda\} + \alpha^2|\beta|^2 \{2\lambda + (m_1^2 - (m_2^*)^2 - m_1(m_2+m_2^*))(1-2\lambda)\} + \\ |\beta|^4 \{\lambda + m_1 m_2^*(1-2\lambda)\} \end{bmatrix}$$

$$+ |1\rangle\langle 1| \ (1/2) \begin{bmatrix} \alpha^4 \{(m_1^2 + 2|m_2|^2)(1-2\lambda) + 3\lambda\} + \alpha^2|\beta|^2 \{(|m_2|^2 + m_1(m_2+m_2^*) + 3m_1^2)(1-2\lambda) + 6\lambda\} + \\ |\beta|^4 \{|m_2|^2(1-2\lambda) + 3\lambda\} \end{bmatrix} \quad (28)$$

The fidelity of the qubit in mode 1 is given by

$$F_3 = \langle\psi|\rho'_1|\psi\rangle \to 3/4 - \alpha^2/2 + \alpha(\beta+\beta^*)/2\sqrt{2} \qquad \text{for } \lambda \to 1/2 \quad (29)$$

If $\beta$ is real, then the average fidelity of this mode is

$$\overline{F}_3 = \int_0^1 F_3(\alpha^2)\, d\alpha^2 \to 1/2 + \pi/8\sqrt{2} = 0.77 \ (approx.) \quad (30)$$

The reduced density operator describing the state $\rho'_2$ is given by

$$\rho'_2 = |0\rangle\langle 0| \ (1/2) \begin{bmatrix} \alpha^4 \{m_1^2(1-2\lambda)+\lambda\} + \alpha^2|\beta|^2 \{(3|m_2|^2 + m_1(m_2+m_2^*) + m_1^2)(1-2\lambda) + 2\lambda\} + \\ |\beta|^4 \{(|m_2|^2 + 2m_1^2)(1-2\lambda) + \lambda\} \end{bmatrix}$$

$$+ |0\rangle\langle 1| \ (1/\sqrt{2}) \begin{bmatrix} \alpha^4 \{m_1 m_2^*(1-2\lambda) - \lambda\} - \alpha^2|\beta|^2 \{(m_1^2 + m_2^2 + m_1(m_2^* - m_2))(1-2\lambda) + 2\lambda\} - \\ |\beta|^4 \{\lambda + m_1 m_2(1-2\lambda)\} \end{bmatrix}$$

$$+ |1\rangle\langle 0| \ (1/\sqrt{2}) \begin{bmatrix} \alpha^4 \{m_1 m_2(1-2\lambda) - \lambda\} - \alpha^2|\beta|^2 \{(m_1^2 + (m_2^*)^2 + m_1(m_2 - m_2^*))(1-2\lambda) + 2\lambda\} - \\ |\beta|^4 \{\lambda + m_1 m_2^*(1-2\lambda)\} \end{bmatrix}$$

$$+ |1\rangle\langle 1| \ (1/2) \begin{bmatrix} \alpha^4 \{(m_1^2 + 2|m_2|^2)(1-2\lambda) + 3\lambda\} + \alpha^2|\beta|^2 \{(|m_2|^2 - m_1(m_2+m_2^*) + 3m_1^2)(1-2\lambda) + 6\lambda\} + \\ |\beta|^4 \{|m_2|^2(1-2\lambda) + 3\lambda\} \end{bmatrix} \quad (31)$$

The fidelity of the qubit in mode 2 is given by

$$F_4 = \langle\Sigma'|\rho'_2|\Sigma'\rangle = (1/2) \begin{bmatrix} R_1(m_1-m_2)(m_1-m_2^*) + R_2(m_1+m_2)(m_1+m_2^*) + \\ R_3(m_1-m_2)(m_1+m_2) + R_4(m_1-m_2^*)(m_1+m_2^*) \end{bmatrix} \quad (32)$$

where

$$R_1 = (1/2)\begin{bmatrix} \alpha^4\{m_1^2(1-2\lambda)+\lambda\}+\alpha^2\beta^2\{(3m_2^2+2m_1m_2+m_1^2)(1-2\lambda)+2\lambda\} \\ +\beta^4\{(m_2^2+2m_1^2)(1-2\lambda)+\lambda\} \end{bmatrix} \qquad (33)$$

$$R_2 = (1/2)\begin{bmatrix} \alpha^4\{(m_1^2+2m_2^2)(1-2\lambda)+3\lambda\}+\alpha^2\beta^2\{(m_2^2-2m_1m_2+3m_1^2)(1-2\lambda)+6\lambda\} \\ +\beta^4\{m_2^2(1-2\lambda)+3\lambda\} \end{bmatrix} \qquad (34)$$

$$R_3 = (1/\sqrt{2})\begin{bmatrix} \alpha^4\{m_1m_2^*(1-2\lambda)-\lambda\}-\alpha^2|\beta|^2\{(m_1^2+m_2^2+m_1(m_2^*-m_2))(1-2\lambda)+2\lambda\}- \\ |\beta|^4\{\lambda+m_1m_2(1-2\lambda)\} \end{bmatrix} \qquad (35)$$

$$R_4 = (1/\sqrt{2})\begin{bmatrix} \alpha^4\{m_1m_2(1-2\lambda)-\lambda\}-\alpha^2|\beta|^2\{(m_1^2+(m_2^*)^2+m_1(m_2-m_2^*))(1-2\lambda)+2\lambda\}- \\ |\beta|^4\{\lambda+m_1m_2^*(1-2\lambda)\} \end{bmatrix} \qquad (36)$$

If $m_1 = m_2 = 1/\sqrt{2}$, then the expression for $F_4$ given in equation (32) reduces to

$$F_4 = R_2 \to 3/4 = 0.75 \quad for \quad \lambda \to 1/2. \qquad (37)$$

Since the machine states are invariant under the unitary transformation T, $\langle A|\rho_3'|A\rangle = \langle A|\rho_3|A\rangle = Y^2$, which is independent of $\alpha^2$.

Hence the deletion machine with transformer becomes a universal deletion machine when the machine parameter $\lambda \to 1/2$ and $m_1 = m_2 = 1/\sqrt{2}$. This universal deletion machine deletes a qubit with fidelity $3/4$ (in the limiting sense), which is the maximum limit for deleting an unknown qubit. In addition, the average fidelity of the qubit in the first mode is found to be 0.77, which is greater than the average fidelity ($\overline{F}_a = 0.66$) obtained by Pati-Braunstein deletion machine.

**Conclusion:** In this paper we define a state dependent, universal and ideal deletion machine and further construct two types of deletion machines. Among these, the first type is the conventional deletion machine, which we generally use, and the second type is a modified version of the first one and hence is named a modified deletion machine. It falls under the category of universal deletion machine. The modified deletion machine deletes a qubit in the second mode with fidelity 0.75 and at the same time, it retains the qubit in the first mode with average fidelity 0.77 (approx.), while the conventional deletion machine deletes a qubit in the second mode with fidelity 0.5 and retains the qubit in the first mode with average fidelity 0.67 (approx.). Hence, the performance of the modified deletion machine is better than that of the conventional deletion machine.

**Acknowledgement:**
I would like to thank Dr. A. K. Pati for useful discussion. I would like to thank Dr. B.S.Choudhury for his support and encouragement during the work. The present work is supported by CSIR project no. F.No.8/3(38)/2003-EMR-1, New Delhi. The support is gratefully acknowledged.